\documentclass[useAMS,usenatbib]{mn2e}

\def\gsim{ \lower .75ex \hbox{$\sim$} \llap{\raise .27ex \hbox{$>$}} }
\def\lsim{ \lower .75ex \hbox{$\sim$} \llap{\raise .27ex \hbox{$<$}} }
\def\Mo{{\rm M_\odot}}

\title[Velocity and spatial biases in CDM subhalo distributions]
{Velocity and spatial biases in CDM subhalo distributions}

\author[J\"urg Diemand, Ben Moore $\&$ Joachim Stadel]
{J\"urg Diemand\thanks{diemand@physik.unizh.ch},
Ben Moore $\&$ Joachim Stadel\\
$$Institute for Theoretical Physics, University of Z\"urich, 
Winterthurerstrasse 190, CH-8057 Z\"urich, Switzerland}

\begin{document}

\pagerange{\pageref{firstpage}--\pageref{lastpage}} \pubyear{2004}

\maketitle

\label{firstpage}

\begin{abstract}

We present a statistical study of substructure within a sample of
$\Lambda$CDM clusters and galaxies simulated with up to 
25 million particles. With thousands of subhalos per
object we can accurately measure their spatial clustering and velocity
distribution functions and compare these with observational data. 
The substructure properties of galactic halos
closely resembles those of galaxy clusters with a small scatter in the
mass and circular velocity functions. The velocity distribution
function is non-Maxwellian and flat topped with a negative kurtosis of
about -0.7.  Within the virial radius the velocity bias
$b=\sigma_{\rm sub}/\sigma_{\rm DM}\sim 1.12 \pm 0.04$, increasing to 
$b > 1.3$ within the halo centers. Slow subhalos are much less common,
due to physical disruption by gravitational tides early in
the merging history. This leads to a spatially anti-biased subhalo
distribution that is well fitted by
a cored isothermal.  Observations of cluster galaxies do not show such 
biases which we interpret as a limitation of pure dark matter 
simulations - we estimate that we are missing half of the halo
population which has been destroyed by physical overmerging.  
High resolution hydrodynamical
simulations are required to study these issues further. If CDM is
correct then the cluster galaxies must survive the tidal field, 
perhaps due to
baryonic inflow during elliptical galaxy formation. Spirals can never
exist near the cluster centers and the elliptical
galaxies there will have little remaining dark matter.  This implies 
that the morphology-density relation is set {\it
before} the cluster forms, rather than a subsequent transformation
of disks to S0's by virtue of the cluster environment.

\end{abstract}


\section{Introduction}

Early simulation work that attempted to follow the merging hierarchy
produced a final dark matter structure that was nearly entirely
smooth (\citealt*{White1976}; 
\citealt*{White1987}; \citealt*{Carlberg1994};
\citealt*{Summers1995}; \citealt*{Tormen1997}). The
reason for this behaviour was debated in the literature as being
due to physical or numerical overmerging 
(\citealt*{White1978}; \citealt*{Carlberg1994}; 
\citealt*{vanKampen1995}; \citealt*{Moore1996}). The
development of fast algorithms to accurately integrate the orbits of
millions of particles overcame this problem. The first
halos simulated with sufficient resolution contained of the order a
thousand substructure halos with properties that resembled galaxies
within clusters \citep*{Moore1998}. These simulations took many months
using parallel gravity codes running on hundreds of processors.

Ongoing research in this area has given many interesting results and we 
list some of the main conclusions here 
(\citealt*{Ghigna1998}; \citealt*{Okamoto1999}; \citealt*{Klypin1999a};
\citealt*{Klypin1999b}, \citealt*{Moore1999}; 
\citealt*{Ghigna2000}; \citealt*{Springel2001}; \citealt*{DeLucia2004}):
(i) Subhalos make up a fraction of between 5 and 10\% of the mass of virialised 
halos.
(ii) Halos on all mass scales have similar substructure populations.
(iii) The mass and circular velocity function of subhalos are power laws with 
slopes -1 and -3.
(iv) Velocity bias between the subhalos and smooth dark matter background may be 
significant.
(v) The radial number density profile of subhalos is shallower than the dark 
matter background.
(vi) Subhalos are significantly rounder than field halos
(vii) The orbits of subhalos are close to isotropic with apo:peri approximately 
4:1.
(viii) Subhalos suffer mass loss from tidal stripping which modifies their outer 
density profiles.
(ix) The tidal radii of subhalos decreases with cluster-centric position.
(x) Most of the surviving population of subhalos entered the parent halo late.

Several of these statements remain controversial and further work is
necessary to clarify certain issues. In this paper we re-address
conclusions (i)-(v) and attempt to answer some of the remaining
questions, including: What is the scatter in the mass and circular
velocity distributions? Is there a positive or negative velocity bias
and if so what is its origin? \citet*{Ghigna2000}; \citet*{Colin2000}
claim a positive velocity bias whilst \citet*{Springel2001} report a
negative velocity bias. Have we converged in the properties of
subhalos, including their radial distribution and mass functions?
The inner regions ($r<$ 0.2 $r_{virial}$) of clusters and galaxy dark
matter simulations are nearly smooth but is numerical overmerging
still occurring in these very high density regions? Does the spatial
distribution of galaxies in clusters resemble that of the subhalos in
simulations? On galaxy scales the observed 
distribution of satellites is more concentrated than the simulations. 
Theory can be reconciled with the observations if it is assumed that
the visible satellites are a biased subset of the total population 
(\citealt*{Taylor2003}; \citealt*{Kravtsov2004}).
On cluster scales we do not expect to
find ``dark galaxy halos'' therefore it is interesting to compare
the observed distribution of galaxies with the distribution of substructure.

In this paper we analyse a sample of six high 
resolution simulations of clusters containing
between 5 and 25 million particles integrated with high force accuracy.
We compare the mass functions with a sample of galactic mass halos with
slightly lower resolution.
These new simulations are presented in Section 2 and the general
properties of the subhalos are given in Section 3.

\section{Numerical Experiments}\label{Sim}

Table \ref{t1} gives an overview of the simulations we present in this paper.
With up to $25\times 10^6$ particles inside the virial radius of one cluster
they are among the highest resolution $\Lambda$CDM
simulations performed so far. They represent a major investment of computing 
time, the largest run was completed in about 5 $10^5$ CPU hours on the zBox 
supercomputer
\footnote{http://www-theorie.physik.unizh.ch/$\sim$stadel/zBox/}.

\begin{table*}
\centering
\begin{minipage}{140mm}
\caption{Parameters of resimulated clusters. 
The last four columns give properties of all subhalos with at least
32 bound particles, their number, bound mass fraction, the 
radius of the innermost subhalo and the
velocity bias $b=\sigma_{\rm sub}/\sigma_{\rm DM}$.
In clusters $A9$ and $C9$ these structures are the cores of 
massive clusters that are about to merge with the main cluster at $z=0$.}
\label{t1}
\begin{tabular}{l|c|c|c|c|c|c|c|c|c|c}
\hline
Run&$\epsilon_0$&$N_{\rm virial}$&
$M_{\rm virial}$&$r_{\rm virial}$&$v_{\rm v,max}$&$r_{\rm vc,max}$&
$n_{\rm halo}$&$\frac{\Sigma m_{\rm halo}}{M_{\rm virial}}$&$r_{\rm sub,min}$&$b$ \\
 &[kpc]& &$10^{15} [\Mo]$&[kpc]&[km s$^1$]&[kpc]
&&&[kpc]& \\
 \hline
 $A9$& 2.4 & 24'987'606 & $1.3 \times 10^{15}$ & 2850 & 1428 & 1853 
&5114&0.07&126*&1.10 \\
 $B9$ & 4.8 & 11'400'727 & $5.9\times 10^{14}$ & 2166 & 1120 & 1321 
&1940&0.12&162&1.12 \\
 $C9$ & 2.4 & 9'729'082 & $5.0\times 10^{14}$ & 2055 & 1090 & 904 
&1576&0.11&77*& 1.15 \\ 
 \\
 $D3h$ & 1.8 & 205'061& $2.8\times 10^{14}$ & 1704 & 944 & 834 
&36&0.03& 260& 1.11 \\
 $D6h$ & 1.8 & 1'756'313 & $3.1\times 10^{14}$ & 1743 & 975 & 784 
&307&0.04& 136& 1.11 \\
 $D6$ &3.6& 1'776'849 & $3.1\times 10^{14}$ & 1749 & 981 & 840 
&322&0.05& 227& 1.13 \\
 $D9$ & 2.4 & 6'046'638 & $3.1\times 10^{14}$ & 1752 & 983 & 876 
&929&0.06& 126& 1.11 \\
 $D9lt$& 2.4& 6'036'701 & $3.1\times 10^{14}$ & 1752 & 984 & 841 
&912&0.05&183& 1.11 \\
 $D12$&1.8& 14'066'458 & $3.1\times 10^{14}$ & 1743 & 958 & 645 
&1847&0.06&136& 1.11 \\
 \\ 
 $E9$& 2.4& 5'005'907&  $2.6\times 10^{14}$ & 1647 & 891 & 889 
&829&0.06&172& 1.11 \\
 \\ 
 $F9$& 2.4& 4'567'075 & $2.4\times 10^{14}$ & 1598 & 897 & 655 
&721&0.06&176& 1.08 \\
 $F9cm$&2.4& 4'566'800 & $2.4\times 10^{14}$ & 1598 & 898 & 655 
&661&0.06&127& 1.08 \\
 $F9ft$&2.4& 4'593'407 & $2.4\times 10^{14}$ & 1601 & 905 & 
464 &706&0.06&161& 1.07 \\
 \\
 $G0$&0.27&1'725'907 & $1.01\times 10^{12}$ & 260 & 160 & 52.2 
&144&0.03&16& 1.05 \\
 $G1$&0.27&1'905'113 & $1.12\times 10^{12}$ & 268 & 162 & 51.3 
&189&0.04&20& 1.03 \\
 $G2$&0.27&3'768'008 & $2.21\times 10^{12}$ & 337 & 190 & 94.5 
&462&0.04&21& 1.10 \\
 $G3$&0.27&2'626'202 & $1.54\times 10^{12}$ & 299 & 180 & 45.1 
&314&0.03&28& 1.12 \\
\end{tabular}
\end{minipage}
\end{table*} 

\begin{figure*}
\vskip 6.0 truein
\includegraphics{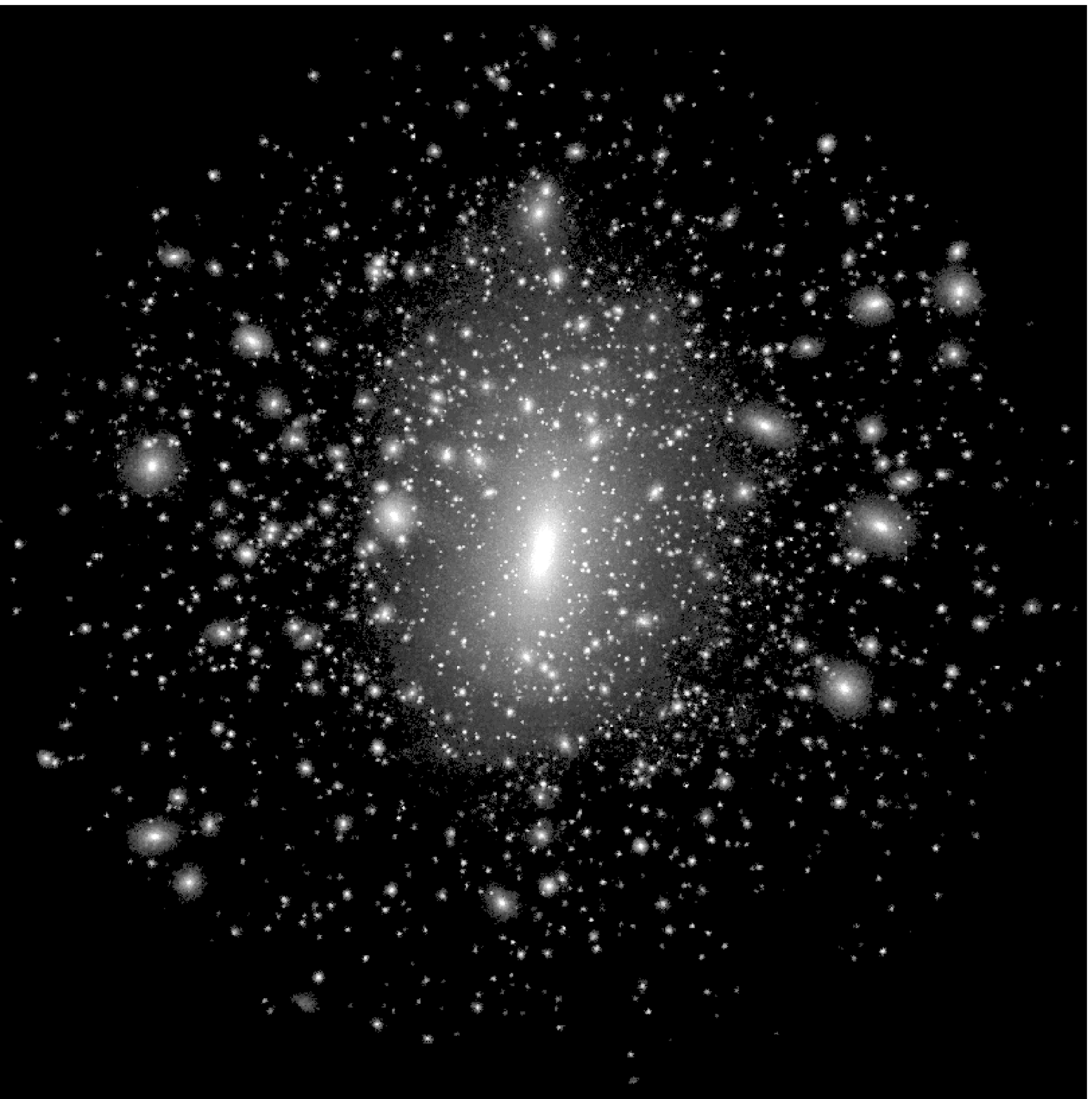}
\caption{\label{pic}Density map for run $D12$ out to the virial radius. This
cluster is prolate with a 3:1 major:minor axis ratio.
Higher resolution color pictures and a mpeg movie of the formation of 
cluster $C9$ can be downloaded from 
http://www-theorie.physik.unizh.ch/$\sim$diemand/clusters/ }
\end{figure*}

\subsection{N-body code and numerical parameters}\label{Nbody}

The simulations were carried out using PKDGRAV written 
by Joachim Stadel and Thomas Quinn \citep{Stadel2001}.
Individual time steps are chosen for each particle proportional to the square 
root of the softening length over the acceleration, 
$\Delta t_i = \eta\sqrt{\epsilon/a_i}$. We use $\eta = 0.2$ for most runs,
only in run $D9lt$ we used larger timesteps $\eta = 0.3$. 
The node-opening angle is set to $\theta = 0.55$ initially and after $z = 2$
to $\theta = 0.7$ to speedup the runs. The code uses a spline 
softening length $\epsilon$, forces are completely Newtonian at $2\epsilon$.
In table \ref{t1} $\epsilon_0$ is the softening length at  $z=0$, 
$\epsilon_{\rm max}$ is the maximal softening in comoving coordinates.
In most runs the softening is constant in physical coordinates from $z=9$ to 
present and constant in comoving coordinates before,
i.e. $\epsilon_{\rm max} = 10 \epsilon_0$. In runs $C9$ and $F9cm$ the softening 
is constant in comoving coordinates for the entire run, in run $F9ft$ the
softening has a constant physical length for the entire run.
 
\subsection{Initial conditions and cosmological parameters}\label{IC}

We use a $\Lambda$CDM cosmological model with parameters from the first year 
WMAP results:
$\Omega_{\Lambda} = 0.732$, $\Omega_m = 0.268$, $\sigma_8 = 0.9$, 
\citep{Spergel2003}.
The initial conditions are generated with the GRAFIC2 package 
\citep{Bertschinger2001}.
Six clusters were selected from a parent simulation and resimulated with much 
higher mass and force resolution, details about the selection and the refinement are 
given in \citet{Diemand2004b}. We label the six cluster 
(ordered by their mass) with letters $A$ to $F$ 
and with a number that gives the refinement factor in length
relative to the $300^3$ in (300 Mpc)$^3$ parent simulation, e.g. 'D12' is the fourth
most massive cluster in our sample, and the mass resolution corresponds to
$(12 \times 300)^3$ particles in a 300 Mpc cube simulation.

We also present results from four medium resolution 
galaxy mass halos which we label G0, G1, G2 and G3.
These halos contain 2-4 million particles within their virial radii. 
The parent simulation is a $90$ Mpc cube resolved with 
$300^3$ particles initially. The four galaxies all lie within a volume of about 
1000 cubic Mpc (at $z=0$) which was refined by a factor of 12 in length to 
reach the resolution given in Table \ref{t1}. 

\subsection{Substructure Identification}\label{si}

Within the virial radius of the high resolution CDM simulations
we can resolve thousands of substructure halos, i.e. self-bound over-dense
clusters of particles (See Figure 1). 
They span a wide range in mass, from the resolution limit of a few tens
of particles up to few percent of the cluster mass, i.e. from 
$10^8 \Mo$ to $10^{13} \Mo$. Some of the subhalos 
even contain their own substructure. Therefore robust identification of 
subhalos a very difficult task, there is no general, parameter free
method that is able to extract the entire hierarchy of halos.

We identify subhalos with SKID \citep{Stadel2001} and with
a new parallel adaptive Friends of Friends ('FoF', see \citet*{Davis1985}) 
group finder ('AdFoF'). SKID calculates local densities
using an SPH kernel, then particles are moved along the density gradient
until they oscillate around a point (i.e. move less than some length $l$).
Then they are linked together using FoF with this $l$ as a linking length. 
AdFoF first calculates the background density of the cluster $\rho_{\rm BG}$ 
using spherical bins. The linking lengths for the particles are set to
$b = (\Delta \rho_{\rm BG} / m_p)^{-1/3}$, $m_p$ is the particle mass, $\Delta = 
5$
is the density contrast, the only free parameter of this method. Two nearby 
particles can now have different linking lengths, 
they are considered as friends if one of them 
considers the other one his friend, i.e. 
the maximum of the two linking lengths is used.
Both the SKID and the AdFoF groups are
checked for self-boundness and unbound particles are removed 
with the same iterative procedure. 

We compared SKID results (using $l = 1.5 \epsilon_0$, $l = 4
\epsilon_0$ and $l = 10 \epsilon_0$) with the AdFof results and we
also visually compared the identified groups with the density map of
the cluster: SKID with $l = 4 \epsilon_0$ adequately identifies the
smallest subhalos and the centers of the largest subhalos. For the
latter the calculated bound mass is underestimated.  Using $l = 10
\epsilon_0$ can be cure this, but then some of the small subhalos are
missed.  The AdFoF has the advantage that in principle it links
together all particles in regions with a density contrast of $\Delta$
against the background density. With $\Delta = 5$ AdFof finds the same
groups as SKID, but the current version using the spherically averaged
density for the background also finds some spurious groups since the
background isodensity surfaces have triaxial shape in a CDM
cluster. For example, particles on the long axis of a prolate halo can
be linked together, since their density is higher than the spherical
average. The subhalo catalogues we analyse in this paper are generated
in two steps: First we use SKID with $l = 4 \epsilon_0$, this gives a
complete catalogue of all the subhalo centers and also the correct
subhalo properties for the smaller objects. Then we run AdFoF with
$\Delta = 5$ and combine the resulting substructure catalogue with the
SKID output to obtain the correct subhalo properties also for the
larger objects: if AdFoF found a subhalo at the same position as SKID,
the properties from the catalogue where this halo has a larger bound
mass are used. The mass fraction bound to subhalos with N $\ge32$
(the cluster centre is not considered a subhalo)
is given in Table \ref{t1}. Using the AdFoF
or the SKID $l = 4 \epsilon_0$ catalogue alone gives about 20 
percent smaller values. Using SKID with $l = 1.5 \epsilon_0$ 
underestimates the masses of the biggest subhalos which
dominate the bound mass fraction, and the results are as much as 
a factor of two below the quoted values.

To check for systematic errors in the substructure catalogue
constructed in this way, we confirmed that the substructure mass
function and the number density profile of one cluster ($D9$) 
remains the same when we construct the substructure catalogue in two
alternative ways: The first alternative catalogue was constructed by
combining three SKID outputs with $l = 1.5, 4$ and $10 \epsilon_0$ as
in \citet{Ghigna2000}, the second alternative was the combination of two
SKID outputs with $l = 1.5$ and $4 \epsilon_0$ and a one $\Delta = 5$
AdFoF output.  We found that the $l = 1.5 \epsilon_0$ SKID does not
find additional structure, the $l = 4 \epsilon_0$ contains all the
small subhalos down to the minimum number of 10. By comparing the
final halo catalogue of cluster D12 to regions of the density map of
this cluster (Figure \ref{pic}) we checked that no subhalos were
missed and that no non-existent halos were included.

\section{Cluster Substructure}\label{Substructure}

We identified subhalos within the virial radii of our six clusters at
redshift zero, the algorithms used are described in section
\ref{si}. At the highest resolution we found over 5000 subhalos 
($\ge$ 32 particles) inside the virial radius of the most 
massive cluster.

\subsection{Spatial Antibias and Convergence Tests}\label{Conv}

\begin{figure*}
\vskip 6.0 truein
\includegraphics{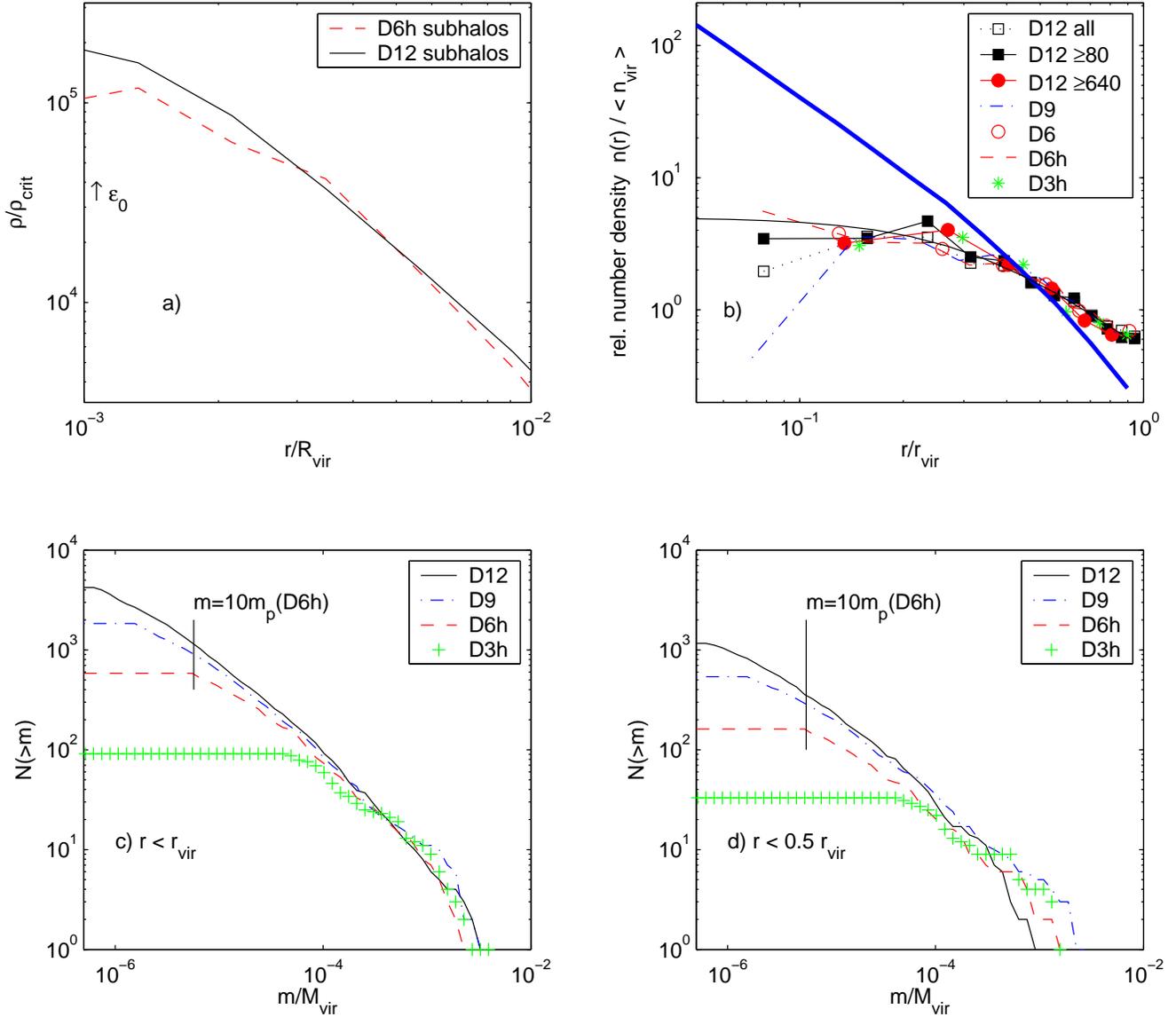}
\caption{\label{subConv.eps}Substructure properties at different mass
resolutions: (a) Average density profiles of 5 subhalos with
masses close to $2.9\times 10^{-4} M_{\rm virial}$ resolved with about
4,000 particles in run D12 and 500 in run D6h. (b) Relative
number density of subhalos with different mass and force resolution,
fitted by an isothermal profile with a core (\ref{nd})
The thick line is the density profile of the DM
particles.  (c) Cumulative mass functions
of substructure within $r_{\rm virial}$ including halos down to 10 $m_p$.
(d) Inner cumulative mass functions, same as (c) but only
including halos within 0.5 $r_{\rm virial}$.}

\end{figure*}

In this section we study the convergence of substructure properties,
including density profiles, cumulative mass functions
and relative number density profiles (Figure \ref{subConv.eps}). First
we check if these properties change with varying force and time
resolution, i.e we compare $D6$ and $D6h$; $D9$ and $D9lt$; $F9$,
$F9cm$ and $F9ft$. The only slight difference we found is in the relative
number density profile: the better force resolution in $D6h$ leads to
a few more surviving substructures near the center (4 subhalos within
10 percent of the virial radius), 
run $D6$ has no subhalos within the same radius. 
Therefore the original numerical
overmerging problem \citep*{Moore1996}
due to insufficient force resolution is not the limiting factor anymore,
except near the center of the halos ($r < 0.1 r_{\rm virial}$).

The amount of substructure that our simulations can resolve is mostly limited by
mass resolution. Subhalos have very high phase space densities,
i.e. relatively short relaxation times. Numerical 
two body relaxation due to finite mass resolution heats up their cores
and makes them less dense \citep{Diemand2004a}. The difference in
central density is about a factor of two between subhalos resolved
with 500 and 4000 particles (see panel (a) of Figure
\ref{subConv.eps} and also
\cite{Kaz2004} where subhalo profiles from clusters
$D6$, $D9$ and $D12$ and their evolution are presented).  
Subhalos with even less particles show this effect more strongly
and have much shallower density profiles. These are less
resistant against tidal stripping and total disruption \citep{Moore1996}.

Figure \ref{subConv.eps} shows substructure properties of the same
cluster, $D$, simulated at different mass resolutions with $N_{\rm
virial}$ = 205k, 1.7M, 6M and 14M.  
Panels (c) and (d) of Figure \ref{subConv.eps} 
show the cumulative mass function including
all subhalos with more than 10 particles. Resolution clearly
affects the numbers of subhalos at the limiting mass of
10 particle masses ($m_p$), however the amount of surviving substructure 
converges at a mass of about 100 $m_p$ for the D6h run. In analogy with
the convergence in density profiles (see  \citealt{Diemand2004b} and 
references therein) we do not expect that this number is
valid for a large range of mass resolutions and it is possible that
the high resolution mass functions are only complete above a mass of 
a few hundred particle masses, especially in the inner region.
We usually include all subhalos with at least
32 bound particles for the analysis presented in this paper, and we will 
always show how the results depend on this minimal number of particles
(in most cases the influence is small).
 
Panel (b) of Figure \ref{subConv.eps} shows the number density of
subhalos in spherical bins relative to the number density within the
virial radius $<n_{virial}> = N_{sub}/V_{\rm virial}$. The first bin is centered
on the innermost subhalo (the cluster center is not considered as a
subhalo), so the first data point also gives the radius $r_{\rm min}$
of the subhalo closest to the center. The size of each bin is set to
$r_{\rm min}$, so the first bin starts at $r_{\rm min}/2$ and ends at
$1.5 r_{\rm min}$.  Tidal disruption is most effective near the
cluster center which leads to an
antibias in the density profile of substructure 
relative to the smooth background. This implies 
that if galaxies are associated with the subhalos, they do not trace
the matter distribution of a cluster. 
Is this antibias real or just an effect of finite resolution? Runs $D6h$, $D9$
and $D12$ have very similar relative number density profiles. If one
only considers groups above the 10 particle limit of $D6h$
(i. e. above 80 $m_p$ in run $D12$), run $D12$ resolves about twice as
many halos as $D6h$ (920 against 582, at the vertical line in
Panel c) and it is interesting to see
where these halos lie. They are not significantly more centrally
concentrated, they have a very similar radial distribution as the
halos that survived in run $D6h$. Even the subhalo distribution of 
all subhalos in $D3h$ (N$\ge$ 10) is very similar to the one of 
the subhalos in $D12$ in the same mass range (N $\ge$ 640) which are
resolved with 64 times more particles. If the convergence scale depends
only mildly on N, for example $r_{\rm converged} \propto N^{-1/3}$ 
as in the case of the density profiles (see  \citealt{Diemand2004b} and 
references therein), the wide range of resolutions
presented here gives for the first time a robust confirmation of 
convergence in the radial distribution of subhalos.
So the antibias in number density does not depend on the 
numerical resolution, but the higher resolution runs 
allow to measure the number density profiles closer to the center.

The relative number density of subhalos 
can be approximated by an isothermal profile with a core
shown by the thin solid line in panel (b) of Figure \ref{subConv.eps}
\begin{equation}  \label{nd}
n(r) = 2 n_{\rm H}\left(1 + (r/r_{\rm H})^2 \right)^{-1} \, ,
\end{equation}
where $n_{\rm H}$ is the relative number density at 
a subhalos scale radius $r_{\rm H}$. The average 
core radius of the distribution of cluster subhalos is
$r_{\rm H} \simeq 0.37$ $ r_{\rm virial} 
\simeq 2/3$ $r_{\rm vc,max}$, where $r_{\rm vc,max}$
is the radius where the circular velocity has its maximum,
see Table (\ref{t1}).

\subsection{Substructure abundance}\label{abundance}

\begin{figure*}
\vskip 6.0 truein
\includegraphics{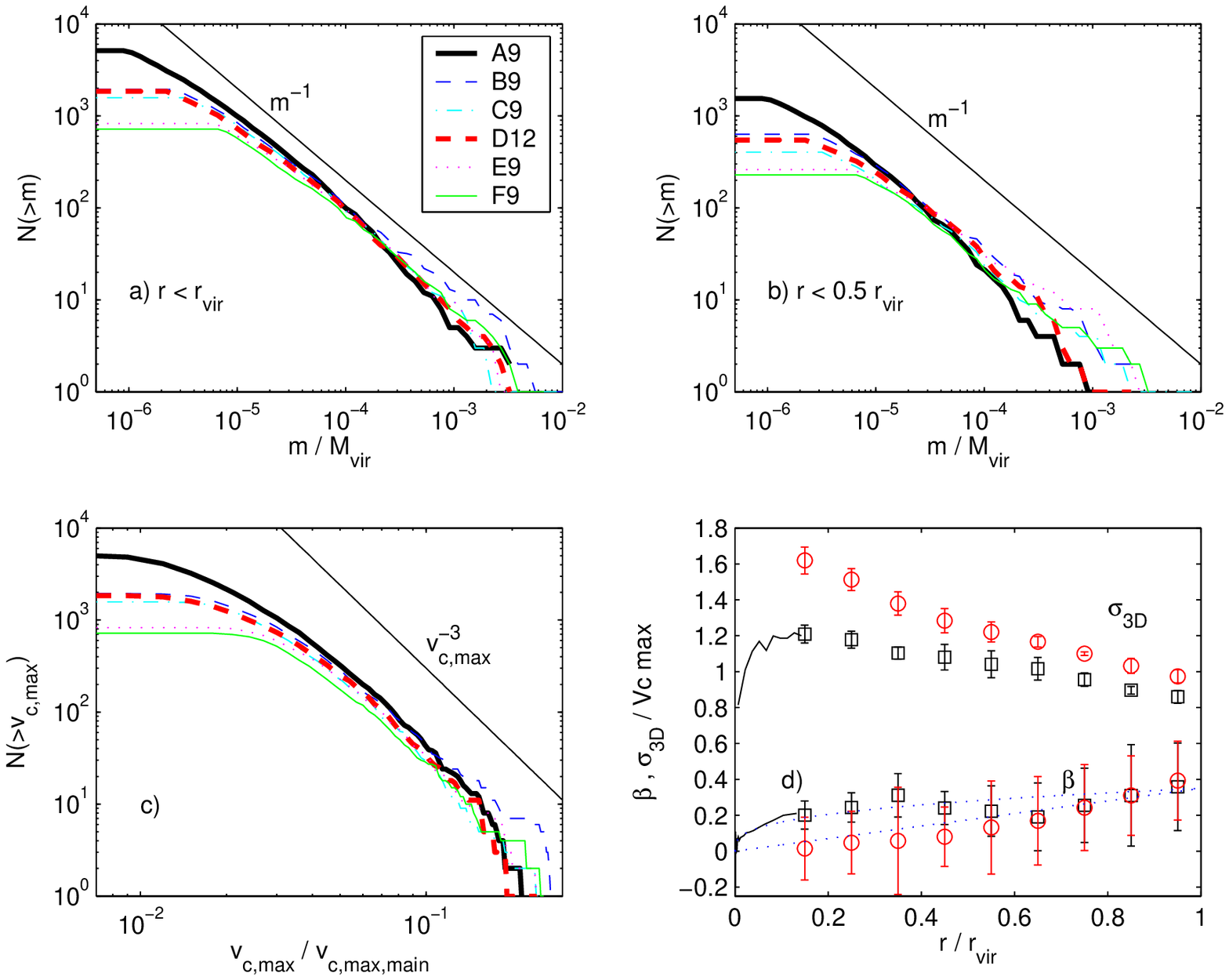}
\caption{\label{subHR.eps} Substructure properties of the six clusters.
Only halos with at least 32 bound particles are considered.
(a) Cumulative mass functions of substructure within $r_{\rm
virial}$. (b) Inner cumulative mass functions, including halos
within 0.5 $r_{\rm virial}$. (c) Cumulative number of subhalos as
a function of their circular velocity. (d) 3D
velocity dispersion of sub halos (circles) and dark matter background
(squares) as a function of radius. Averages over all 6 cluster
profiles, normalised to the maximum circular velocity.  
Error bars show the scatter between the clusters. Poisson errors due to
small number of subhalos per bin are smaller than $0.05$ and are not
included. The average of the anisotropy parameter 
$\beta = 1 - 0.5 \sigma_t^2/\sigma_r^2$ 
is also plotted for the subhalos (circles) and the particles (squares). 
The particles are on slightly more radial orbits than the subhalos.
The dotted lines are fitting functions, see text for details.}
\end{figure*}

Figure \ref{subHR.eps} shows the cumulative substructure 
mass functions
and inner mass functions of the
six clusters which are all well approximated by a simple power law
$m^{-1}$. Here we include subhalos with a minimum of 32 particles, we
found in the last section that the subhalo catalogues are complete
only above a mass corresponding to about 100 particles. The apparent
flattening of the slope towards this mass is due to finite resolution
and does not indicate a shallower power law at lower masses. This can
also be seen from the fact that around $m = 10^{-5} M_{\rm virial}$ the
larger halos and run $D12$ (i.e. those with better relative mass resolution) 
have
steeper slopes. If hierarchical merging should produce subhalo mass
functions that do not depend on the mass of the parent halo (as shown
in \cite{Moore1999}, see also Section \ref{galaxies} of this paper)
the natural outcome is an $m^{-1}$ power law: If one simply adds two
equal halos the amount of substructure above any fixed absolute mass
doubles, the remnant has now twice the mass and it only has the same
amount of substructure at a fixed relative mass if the mass function
of the progenitors was $m^{-1}$.
The mass function of isolated field halos is also close to a power law
of slope $m^{-1}$ (e.g. \citealt*{Jenkins2001}; \citealt*{Reed2003}). 
Thus tidal stripping acts to lose mass in such a
way that the overall mass function slope does not change. 
The conspiracy is such that stripped halos move down the
M versus $v_{\rm c,max}$ plane such that they follow the line for field halos
\citep{Ghigna1998}.

The cumulative substructure velocity functions (see Figure
\ref{subHR.eps}, Panel (c)) 
gives the number of subhalos with maximum circular
velocities above a given value. The virial theorem $v_{\rm c,max}^2
\propto M_{\rm halo}/R \propto M_{\rm halo}/M_{\rm halo}^{1/3}$ leads
to a simple scaling $M_{\rm halo} \propto v_{\rm c,max}^3$ for field
halos. This relation is also a good approximation for subhalos, even
if they lost most of their mass due to tidal stripping
(\citealt{Ghigna1998}; \citealt{Kravtsov2004}). Since the cumulative mass
function goes like $m^{-1}$, we expect the cumulative mass functions
to follow a $v_{\rm c,max}^{-3}$ power law. This is true in a wide
range of velocities. Towards the resolution limit the velocity
functions also become shallower, but this is due to the same numerical
effect as in the case of the mass functions.
The scatter in the substructure abundance is large at the high mass end
(a factor three) where the mass functions depend on
a small number of massive objects. At
intermediate and small subhalo masses 
($<10^{-4}M_{vir}$) the scatter is within a factor of 1.7.

\subsection{Subhalo velocity distribution}

\subsubsection{Velocity bias}

Figure \ref{subHR.eps}, panel (d) shows the 3D velocity 
dispersion of the smooth particle
background and subhalos. We measured the dispersion
profile for each individual cluster, 
then we averaged the values in each 
bin over all six clusters. The subhalos dispersions
are not weighted by mass, each subhalo has equal weight.
In a radial range from 0.1 $r_{\rm virial}$ to 0.4
$r_{\rm virial}$ the substructure halos have a higher 3D velocity
dispersion than the background: $b = \sigma_{sub}/\sigma_{DM}$ is $b =
1.25 \pm 0.08$. The velocity bias of all subhalos within the virial
radius $b = \sigma_{sub}/\sigma_{DM}$ is $b = 1.11 \pm 0.04$.  The
plotted and quoted errors are the scatter in our sample of six
clusters and they are much larger than the Poisson noise in the
estimated values of $\sigma_{sub}$.

A negative velocity bias was first considered by \cite{Carlberg1989} as a
possible way of reconciling low cluster masses with a high
matter density universe. 
Hints for positive bias ($b>1$) were found 
by \cite*{Ghigna1998} and also \citet*{Colin2000} who combined 12 clusters 
containing 
33 - 246 resolved subhalos
to obtain a sufficiently large subhalo sample.
The first simulation with sufficient resolution
(about 5 million particles within the virial radius) 
to construct a reliable subhalo velocity dispersion profile
from one object was analysed in
\citet{Ghigna2000}. They found 
$b = 1.2 - 1.3$ in their innermost bin,
which goes from 0 to 0.25 $r_{\rm virial}$, and a small ($< 1.10$)
positive bias for the entire cluster. 

The bias is independent of subhalo mass, for example including only halos
above $5\times 10^{-5} M_{\rm virial}$ (979 subhalos or about 8 percent of the
subhalos with $N\ge32$) also gives $b = 1.11 \pm 0.04$. 
And for halos above $10^{-4}
M_{\rm virial}$ (only 474 halos or 4 percent) $b = 1.10\pm 0.05$.
The velocity bias does not depend on resolution: In the radial
range from 0.1 $r_{\rm virial}$ to 0.4 $r_{\rm virial}$ the values lie
within $b = 1.16$ and $b = 1.25$ for all simulations of cluster $D$
and there is no clear trend with resolution.

\subsubsection{Anisotropy of subhalo velocities}

In the radial and tangential velocity dispersions 
the bias is very similar 
as in the three dimensional dispersion.
This can also be seen from the anisotropy parameter
$\beta =1-0.5 \sigma_t^2 /  \sigma_r^2$, (Panel (d) in Figure \ref{subHR.eps}):
The anisotropy is very similar for subhalos and background 
particles, only in the inner region
the subhalo velocities are slightly more isotropic than those
of the particle background.
From $r=0$ to $r_{\rm virial}$ the anisotropy $\beta$ grows roughly 
linear with radius: $\beta \simeq 0.35 r$. For the average
particle anisotropy $\beta \simeq 0.35 r^{1/3}$ seems to fit the data better.

\subsubsection{Subhalo dynamics}

Here we investigate if the spatial and velocity distribution can be 
a steady-state solution of the collisionless Boltzmann equation (CBE)
or if a supply of infalling structures is needed to maintain the
state of the system observed at $z=0$. We neglect the small anisotropy
and assume spherical symmetry, then the integral of the second
moment of the CBE, the Jeans Equation \citep{Binney1987}, reads 
\begin{equation}\label{jeans}
\rho_{\rm sub}(r)\sigma_{\rm r,sub}^2(r) = \int_r^c \rho_{\rm sub}(r) \frac{GM(r)}{r^2} dr  
\end{equation}
where $c$ gives the size of the system, 
$\rho_{\rm sub}$ and $\sigma_{\rm r,sub}$ are the density and the one-
dimensional
dispersion of the subhalos and 
$M(r)$ is the cumulative {\it total} mass. A similar equation for the
dark matter background is obtained by using density and dispersion
of the dark matter instead.

\begin{figure}
\vskip 3.0 truein
\includegraphics{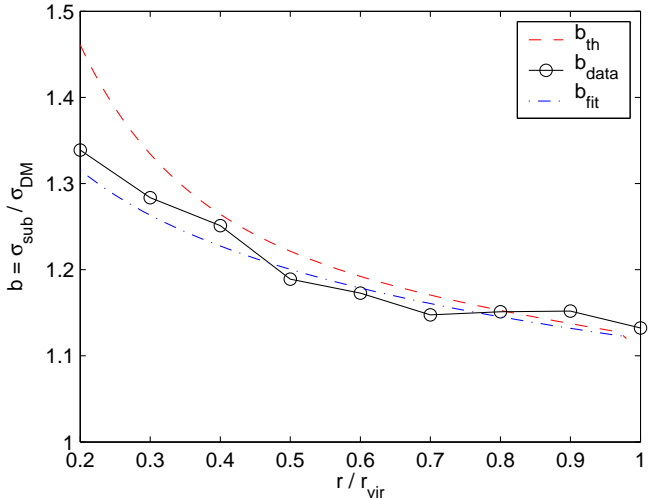}
\caption{\label{jeansBias.eps}
Velocity bias profile. Circles give the average bias of the six clusters.
The dashed line is the bias calculated from the Jeans equation (\ref{bth})
using the different density profiles of subhalos and background particles
and assuming that the two are in dynamical equilibrium. The dashed-dotted
line gives a simple power law fit $\propto r^{-0.1}$ to the average bias.
} 
\end{figure}

The six clusters can be approximated as NFW profiles \citep{Navarro1996} 
with a mean concentration of about $c_{\rm NFW} = 7$ (see \citealt{Diemand2004b}).
Using this average dark matter density profile the $\sigma_{\rm r,DM}^2(r)$ from 
Equation (\ref{jeans}) fit the measured values (Figure \ref{subHR.eps})
very well.
For the radial density profile of the subhalos we use Equation (\ref{nd}),
with $r_{\rm H} = 2/3$ $r_{\rm vc,max}$, the mean of
$r_{\rm vc,max}$ is about 0.57 $r_{\rm virial}$.
The expected bias is
\begin{equation}\label{bth}
b_{\rm th} = \frac{\sigma_{\rm r,sub}(r)}{\sigma_{\rm r,DM}(r)} =
\left[  \frac{\rho_{\rm DM}(r)}{\rho_{\rm sub}(r)}
\frac{\int_r^c \rho_{\rm sub}(r) \frac{GM(r)}{r^2} dr}
{\int_r^c \rho_{\rm DM}(r) \frac{GM(r)}{r^2} dr} \right] ^{1/2} \, .
\end{equation}
We use a cut off at $c = 2$ $r_{\rm virial}$, at this 
radius the slopes of $\rho_{\rm sub}$ and $\rho_{\rm DM}$ become similar 
and the bias should vanish.
Figure \ref{jeansBias.eps} shows the predicted and measured velocity
bias and simple power law fit to 
the measured average velocity bias:
$b_{\rm fit} = 1.12 \times (r/r_{\rm virial})^{-0.1}$.
$b_{\rm th}$ is very close to
the measured velocity bias, just in the inner region $b_{\rm th}$ is 
too large. This means that the subhalo-background system is close to a 
steady-state equilibrium configuration. 

Therefore we expect the non-equilibrium processes to be subdominant. The
net {\it infall} of subhalos can be quantified from the asymmetry of the radial
velocity distribution of subhalos near the virial radius: The distributions are
symmetric in the inner and outer part of the clusters and there is no net infall of 
subhalos at z=0. Another non-equilibrium process is the {\it disruption} of subhalos. 
The fraction of subhalos that are disrupted is small (see also
Section \ref{origin}), about 0.02 Gyr$^{-1}$ for subhalos
with N $\ge 100$. In the inner 40 percent of the halo the fraction is bigger,
about 0.13 Gyr$^{-1}$. This could be the reason why the steady-state solution 
over-predicts the velocity bias near the center.

\begin{figure}
\vskip 4.5 truein
\includegraphics{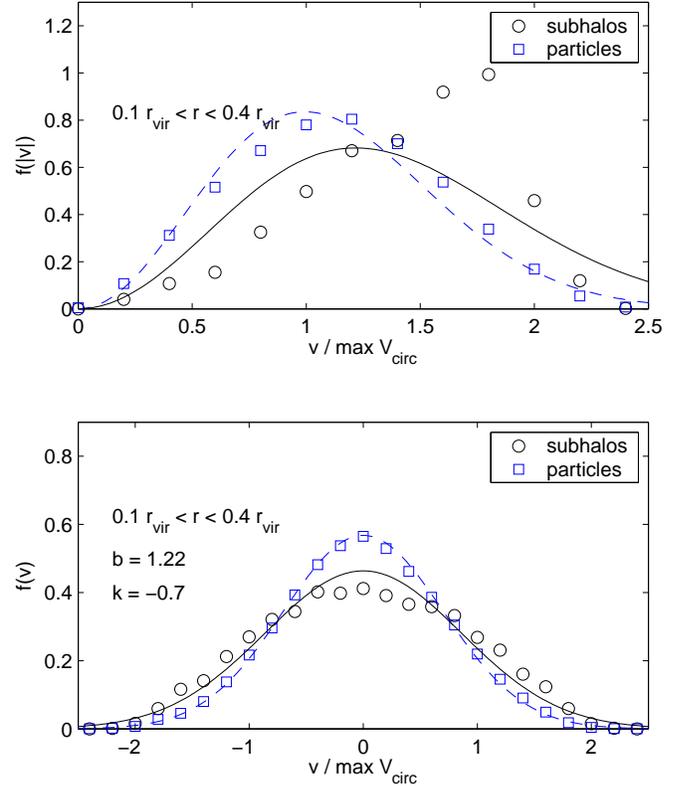}
\caption{\label{innerVdis.eps}Velocity distribution 
of inner subhalos (circles) and particles (squares).
Average of six distributions from different clusters.
Subhalos and particles between $r = 0.1 r_{\rm virial}$ and $0.4 r_{\rm virial}$
are included.
Velocities are normalized to the maximum circular velocity $v_{\rm c,max}$
of each cluster. Solid and dashed lines are 
Maxwellian distributions with the correct second moment.} 
\end{figure}

\begin{figure}
\vskip 4.5 truein
\includegraphics{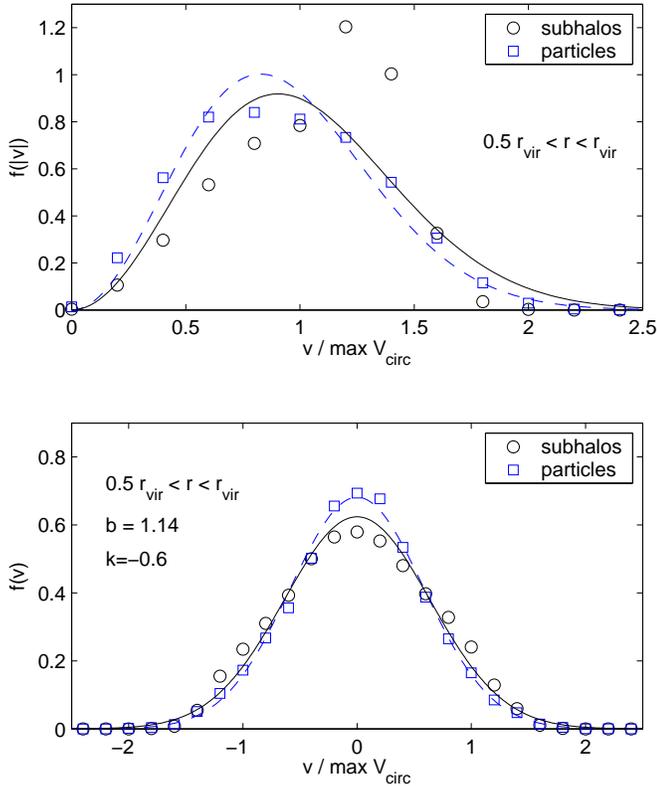}
\caption{\label{outerVdis.eps}Same as Figure \ref{innerVdis.eps}, but including
subhalos and particles between $r = 0.5 r_{\rm virial}$ and $r_{\rm virial}$.} 
\end{figure}

\subsubsection{Higher moments of the velocity distribution}

In the last subsection we found that the second moment of the subhalo
velocity distribution is consistent with a steady-state solution, 
where the subhalos have a spatial antibias. Now we consider
the next higher moments of the velocity distributions of
subhalos and particle background. 
In the radial range where the velocity bias is
large (0.1 $r_{\rm virial}$ to 0.4 $r_{\rm virial}$) the
shapes of these velocity distributions are
very different (Figure \ref{innerVdis.eps}). There are many less
subhalos with small velocities (top panel), also the fraction of
subhalos with low velocity components is smaller for the
particles (bottom panel). While the particle velocity distribution is
close to a Maxwellian, this is not true for the subhalos. The subhalo
velocity histogram is flat-topped, it has smaller fourth moment than
the Maxwell distribution, i.e. a negative kurtosis $k= <v^4> / <v^2>^2
- 3 = -0.7$. We also calculated the first two non-trivial,
even\footnote{ The odd moments are zero for symmetric functions. } 
Gauss-Hermite moments $h4$, $h6$ \citep{Gerhard1993}. In this radial range
(0.1 - 0.4 $r_{\rm virial}$) we get $h_4 = -0.068$ and $h_6 =
0.0013$. The advantage of Gauss-Hermite moments over simple higher
order moments is that they are not very sensitive to the wings of the
distribution. In galaxy clusters these outer parts of the distribution
are hard to determine exactly due to interlopers
\citep{VanDerMarel2000}.

In Figure \ref{outerVdis.eps} we plot the velocity histogram further
out (0.5 $r_{\rm virial}$ to $r_{\rm virial}$). Now the second moments of the
particle and subhalo velocities are much closer ($b=1.10$), but the
shapes of the velocity distributions of subhalos and particles
are still different: $k = -0.60$, $h_4 = -0.031$ and $h_6 =
-0.025$. For all subhalos within $r_{\rm virial}$ we find $b=1.11$, $k =
-0.48$, $h4 = -0.034$ and $h_6 = -0.012$.

Both the inner (Figure \ref{innerVdis.eps}) and outer 
(Figure \ref{outerVdis.eps}) subhalos show an {\it excess of high-velocity
substructures} between $v_{\rm c,max}$ and 1.5 $v_{\rm c,max}$ . Many of these 
high-velocity subhalos are on very radial orbits. When we exclude subhalos
with absolute values of the radial velocity component larger 
than $v_{\rm c,max}$ the excess disappears
and the speed distribution follows the Maxwellian distribution of the background particles
above $v_{\rm c,max}$ . The large fraction of subhalos with very high radial velocities
is also evident in the radial velocity distribution (not shown): both in the inner
and outer part of the clusters the distribution has a very negative kurtosis of $k = -0.9$.
Also note that the radial velocity distributions are symmetric, there is not net infall of 
subhalos at z=0.

The shape parameters depend weakly on the lower mass threshold,
including subhalos above $5\times10^{-5} M_{\rm virial}$ instead of 32
$m_p$ yields: $b=1.11$, $k = -0.44$, $h4 =
-0.016$ and $h_6 = -0.022$. There are 979 subhalos above this
threshold in our six clusters, which is only $979/12027 = 0.039$ of
the $N\ge32$ subhalo sample, 
but this is still enough to determine the shape of
the velocity distribution. All of these subhalos have bound masses of 
more than $1.2\times10^{10}\Mo$.

\subsection{The origin of the subhalo biases}\label{origin}

The physical mechanism that generates the differences in the spatial 
and velocity
distributions of particles and subhalos is most likely the tidal
destruction of subhalos in dense environments. 
The efficency of tidal stripping and tidal
disruption depends mostly on the orbital energy of the subhalos
(\citealt{Ghigna1998}; \citealt{Taffoni2003}; \citealt{Kravtsov2004}).
Therefore it offers a natural explanation for the lack of slow
subhalos; at a fixed radius the orbital energy is 
proportional to the square of the velocity and tidal disruption
could remove a large fraction of the slow subhalos
producing a distribution like the one given in
the top panel of Figure \ref{innerVdis.eps}.

The tidal disruption of subhalos must occur very early in the evolution
of the cluster. \citet{Ghigna2000} are able to identify the 
remnants of 60 to 70 percent of all cluster progenitor halos 
($N\ge100$ at z=3) with subhalos at z=0.  
\footnote{The fraction of subhalos that merge with the central object
(i.e., end up within an assumed radius of  about 0.015 $r_{\rm virial}$)
are always below 5 percent and
can be neglected in this context. But it is an important 
fraction if one considers the most 
massive progenitors only \citep{Ghigna2000}.}. 
From the halos identified at $z=1$ an even larger fraction survives 
(more than 80 percent). For run $D6h$ we performed the same test and 
get very similar numbers. We link progenitor halos with
a halo at $z=0$ if at least four particles of the progenitor 
are bound to the subhalo at $z=0$ and find descendents 
for 83 percent of the progenitor halos identified at z=2.

However a significant fraction of subhalos 
may have been destroyed prior to this epoch. From the halos 
with $N\ge100$ identified
in the high resolution region of run $D6h$ at z=7.2 and z=4.3, we can 
associate only about 60 percent with $z=0$ subhalos. At this early stage
tidal disruption seems to act as a physical selection process
which allows only halos with high enough orbital energies to survive as
todays subhalos. This causes the spatial antibias and the 
positive velocity bias of substructure. 

Note that it is important to have a larger minimum number of bound 
particles in the early subhalo sample ($N\ge100$) than in the final subhalo 
catalogue ($N\ge10$) if one wants to quantify disruption: If we would use 
the same N at both times then we would get a much higher 'disruption rate',
but we would mostly measure the amount of subhalos that were tidally stripped
below this threshold number of bound particles but not necessarily disrupted.
This caveat would have a big 
influence since about half of the considered subhalos have a 
bound mass between N$m_p$ and 2N$m_p$.

\subsection{Comparison with galaxy size halos}\label{galaxies}

The four galaxies in our sample are resolved with 1.7 to 3.8 million
particles, so the relative mass resolution is lower than for the
clusters. However, there is enough resolved substructure to compare its
abundance and the radial distribution to the results from
the cluster runs. We make the comparison with 
cluster $D6h$ which has similar relative mass and force
resolution as the galaxies. We also give the results for the same
cluster with eight times better mass resolution (run $D12$) to get an
impression how the results might change if we also had higher
resolution for the galaxy halos. There may be a hint that the 
galaxies have slightly less substructure than the clusters, but we
need to increase the resolution in the galaxy simulations in order
to verify this result.

Galaxy G2 had a recent major merger at $z \simeq 0.2$, at $z = 0$ this merger
is finished, the core has no more visible signs of dynamical
activity. The concentration of this galaxy is lower $c_{\rm vc,max} =
r_{\rm virial} / r_{\rm vc,max} \simeq 3.6$, probably due to the later
formation in this recent merger. The other three galaxies had no more 
major mergers since at least $z \simeq 0.2$ and 
their $c_{\rm vc,max}$ are between 5 and 6.5.

\subsubsection{Substructure abundance}

\begin{figure*}
\vskip 6.0 truein
\includegraphics{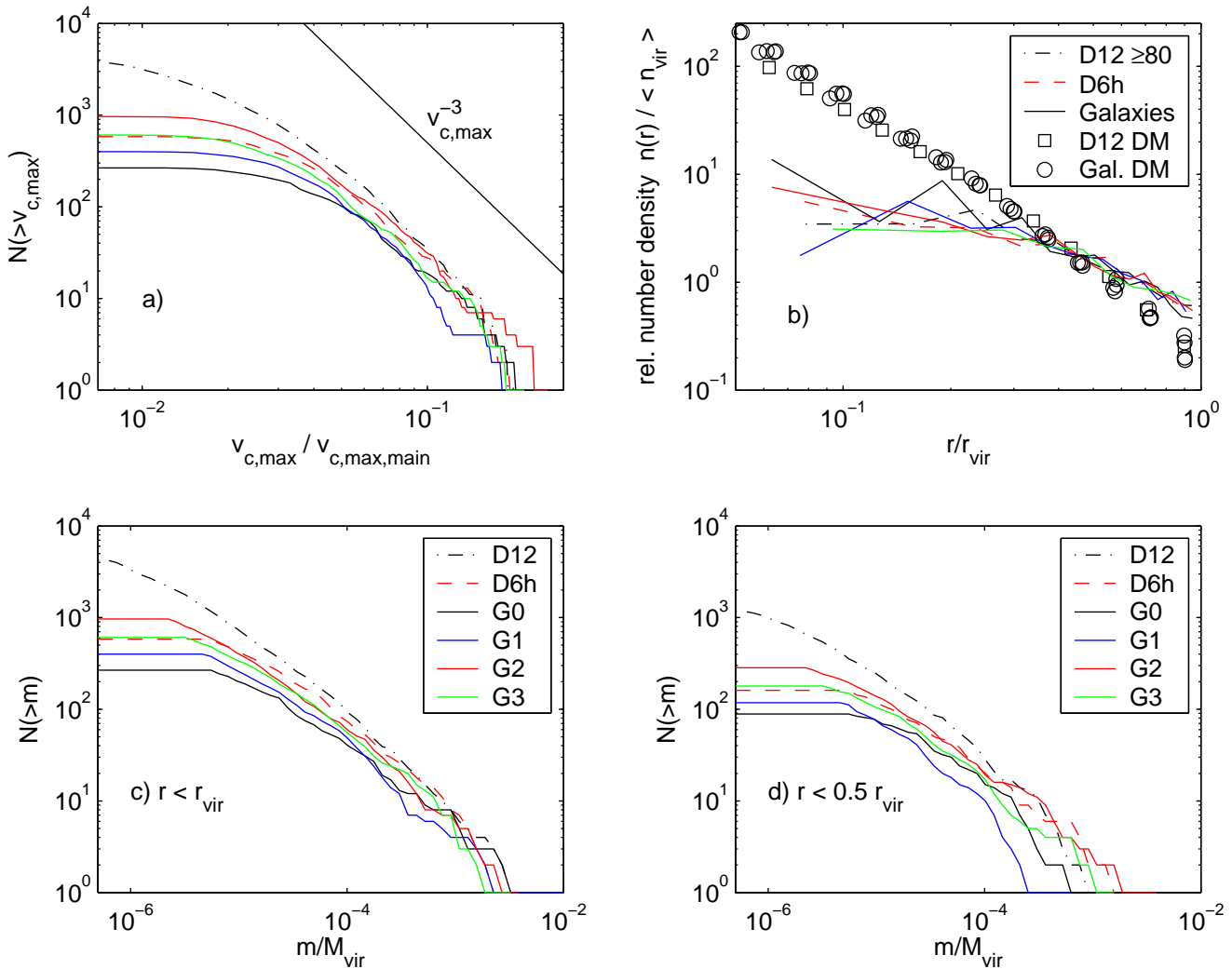}
\caption{\label{subCandG.eps} Substructure properties of four galaxy
halos: Panel a): Cumulative number of subhalos as a function of their
circular velocity.  Panel b): Relative number density of subhalos and
of all DM particles (see Section \ref{Conv} for details).  Panel c):
Cumulative mass functions of substructure within $r_{\rm virial}$.  Panel
d): Inner cumulative mass functions, including halos within 0.5
$r_{\rm virial}$.  All halos with at least 10 bound particles are
included in these plots.  The solid lines show the four galaxies, the
dashed line is a cluster halo at similar resolution and the dashed
dotted line is the same cluster at eight times higher force resolution
for comparison.}
\end{figure*}

Despite the fact that clusters form much later than galaxies in
hierarchical structure formation, they have very similar subhalo mass
function. \cite{Moore1999} showed this by comparing two SCDM halos.
\cite{DeLucia2004} confirmed this recently for several $\Lambda$CDM 
halos, but at a resolution of
less than a million particles inside the virial radius.
Figure \ref{subCandG.eps} shows the subhalo abundance in the four
galaxies and in the cluster $D$.  The velocity functions (Panel a),
mass functions (Panel c) and inner mass functions (Panel d) are all
quite close to those of the reference cluster run $D6h$.

The substructure abundance is largest in galaxy $G2$, it is as high as
in run $D6h$. This halo formed recently in a major merger at $z \simeq
0.2$, which is a typical formation history for cluster size halos
rather than for galaxies. The other three galaxies have about 30
percent less substructure than $G2$ and $D6h$. Therefore the amount of
substructure depends weakly on the mass of the parent halo, but the
difference appears to be comparable to the scatter within parent halos
of a fixed mass.
 
\subsubsection{Radial distribution}

The relative number density profiles (Panel (b) of Figure
\ref{subCandG.eps}) of the galaxy subhalos are more centrally
concentrated than those of cluster subhalos \citep{DeLucia2004}.
Smaller halos have higher concentrations \citep{Navarro1996} and are
therefore more resistant against tidal disruption. However the
subhalo number density also shows a clear antibias with respect
to the dark matter density. 

The density profile that fits the 
cluster subhalos distribution (Equation \ref{nd})
is a good approximation also 
for the galaxy subhalo number density
profile. Now the core radius is a smaller 
fraction of the virial radius 
($r_{\rm H} \simeq 0.14$ $r_{\rm virial}$) because
galaxy subhalos are more centrally concentrated.
Note that $r_{\rm H}$ is again about two thirds of the 
radius where the circular velocity is maximal, 
this is the same fraction as for the cluster subhalos.
Therefore scaled to $r_{\rm vc,max}$ galaxy and cluster subhalos 
number density profiles are the same.

\section{Comparison with observations}

\subsection{Substructure abundance}

\cite{Desai2004} measured galaxy circular velocity function in 34 low-redshift
clusters and found that these functions can be approximated by a power-law
$\propto v_{\rm c,max}^{-2.5}$. In CDM cluster simulations they found a 
logarithmic 
slope of $-3.4\pm 0.8$. Our higher resolution simulations show that these
slopes are rather on the steep side of the given range, Figure (\ref{subHR.eps})
shows that the {\it cumulative} velocity function has a slope of about
$-3$, where we expect the sample to be complete. For the
differential circular velocity function this gives a slope of $-4$,
which is not consistent with the observed slope of $-2.5$.
Accounting for the effects of the baryons could reconcile CDM simulations
with the observations, see e.g. \citet{Springel2001} and \citet{Desai2004}.
Realistic gas-dynamical cluster simulations 
will eventually resolve this issue.

The same problem is more severe when the host halo is a galaxy and not a
cluster. The steep circular velocity function of CDM halos predicts over 100
subhalos with circular velocities above 5 percent of the parent halo circular 
velocity, i.e. above 10 km/s for a Milky Way size halo. Our highest resolution
cluster $A9$ has over 300 subhalos above this velocity. But the number of
Milky Way satellites with $v_{\rm c,max} > 0.05 v_{\rm c,max,parent}$  is only 10 
\citep{Moore1999}. 
Various solutions to this issue have been proposed in the literature 
(e.g. \citealt*{Stoehr2002}, \citealt{Kravtsov2004}).

\subsection{Spatial distribution}

\begin{figure}
\vskip 3.0 truein
\includegraphics{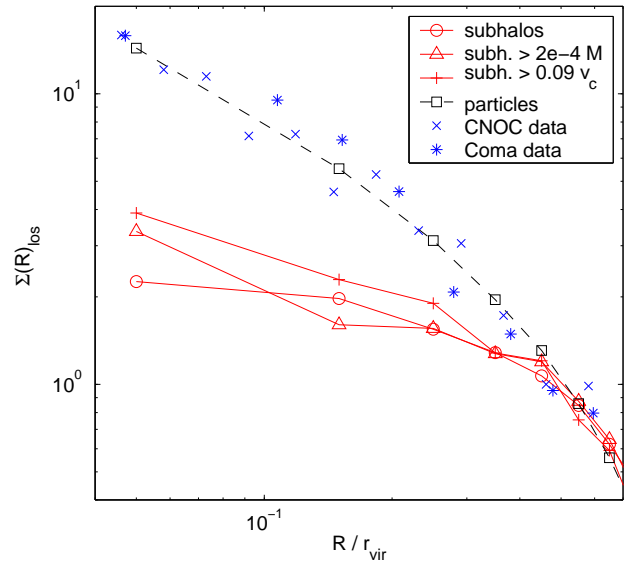}
\caption{\label{surfProj.eps} Projected relative number 
surface density profile of subhalos averaged over the six clusters:
Circles include all (12'023) subhalos with $N \ge 32$,
triangles only halos with $m > 10^{-4} M_{\rm virial}$ and
plus signs only halos with $v_{\rm c,max} > 0.09 v_{\rm c,max,main}$. 
The core of the main halo, the `cD galaxy', is always included in the first bin.
The projected dark matter density is plotted with squares.
Crosses are the data from the CNOC survey \protect\citep{Carlberg1997},
stars are the Coma cluster data from \protect\cite{Lokas2003}. 
We normalise the curves so they match at $r_{\rm virial}$.}
\end{figure}

For comparison with observed spatial and velocity distributions of galaxies 
in clusters we `observe' the six simulated clusters 
along three different line of sights (LOS) (the x, y and z axis) and average 
over these LOS. We then take the sample averages to get mean values and
an estimate of the scatter. The results are shown in Figure \ref{surfProj.eps}
and \ref{biasProj.eps}.

The number surface density is plotted at the midpoints of equal bins
in projected distance from the densest region of the cluster. The
innermost bin starts at $R = 0$ and therefore always one additional
subhalo, i.e. the core is counted as the cD galaxy of the cluster. The
projected number density is flat near the center, just like the 3D
number density in Figure \ref{subConv.eps}. The total sample contains
12'027 subhalos with at least 32 bound particles from the 6 high
resolution clusters. 
In the Coma cluster a number density profile for a comparable
number of galaxies (985) can be measured \citep{Lokas2003}, this
profile (plotted with stars in Figure \ref{surfProj.eps}) 
is steeper than the subhalo profile and
follows rather closely the expected dark matter profile of a CDM cluster.
\cite{Carlberg1997} give the surface density profile of a sample of
galaxies combined from 14 clusters observed in the CNOC cluster
survey. The sample contains 1150 galaxies, including background and
goes out to $2 r_{200}$, i.e.,  per cluster there are about 50
galaxies. Therefore this magnitude limited sample should be comparable
to the most massive 300 subhalos in our sample. 

We selected the subhalos
with $m > 2\times10^{-4} M_{virial}$ and get a sample of 238 halos, their 
surface
density profile is plotted with triangles in Figure \ref{subConv.eps}.
The profile does not change much, just in the innermost bin the values
rise, due to the relative importance of the 'cD galaxy'. 
Selecting subhalos by peak circular velocity 
$v_{\rm c,max} > 0.09$ $ v_{\rm c,max,parent}$ gives a sample of 291 halos
with a similar surface density profile.

The observed number surface density profiles from \citet{Carlberg1997}
and \citet{Lokas2003}
(and also \citet*{Beers1986} and \citet*{Merrifield1989})
are significantly steeper than in the CDM clusters. To
correct the subhalo number density in the inner four bins upwards to
match the observed values one needs to add a number of subhalos
similar to the total number within the virial radius of each cluster,
but preferentially more subhalos closer to the cluster center.
We discuss the implications of this result in the conclusions.

\subsection{Subhalo velocities}

The velocity bias $b \sim 1.12 \pm 0.04$ would lead to dynamical cluster
mass estimates that are about 20 percent too high if cluster galaxies
reside in CDM subhalos. By comparing with cluster mass estimates from 
gravitational lensing it could be noted the dynamical estimated
are too high, but it is very difficult to obtain estimates with small
enough uncertainties with both methods. Such a comparison was performed
by \cite{Cypriano2001}, finding that dynamical masses are 
indeed biased by $1.20 \pm 0.13$ in a sample of 14 clusters, 
but the effect only comes from the massive clusters 
($\sigma_v > 1122$ km/s), which show large mass differences $1.54 \pm 0.19$,
while the smaller clusters show no bias.

\begin{figure*}
\vskip 6.0 truein
\includegraphics{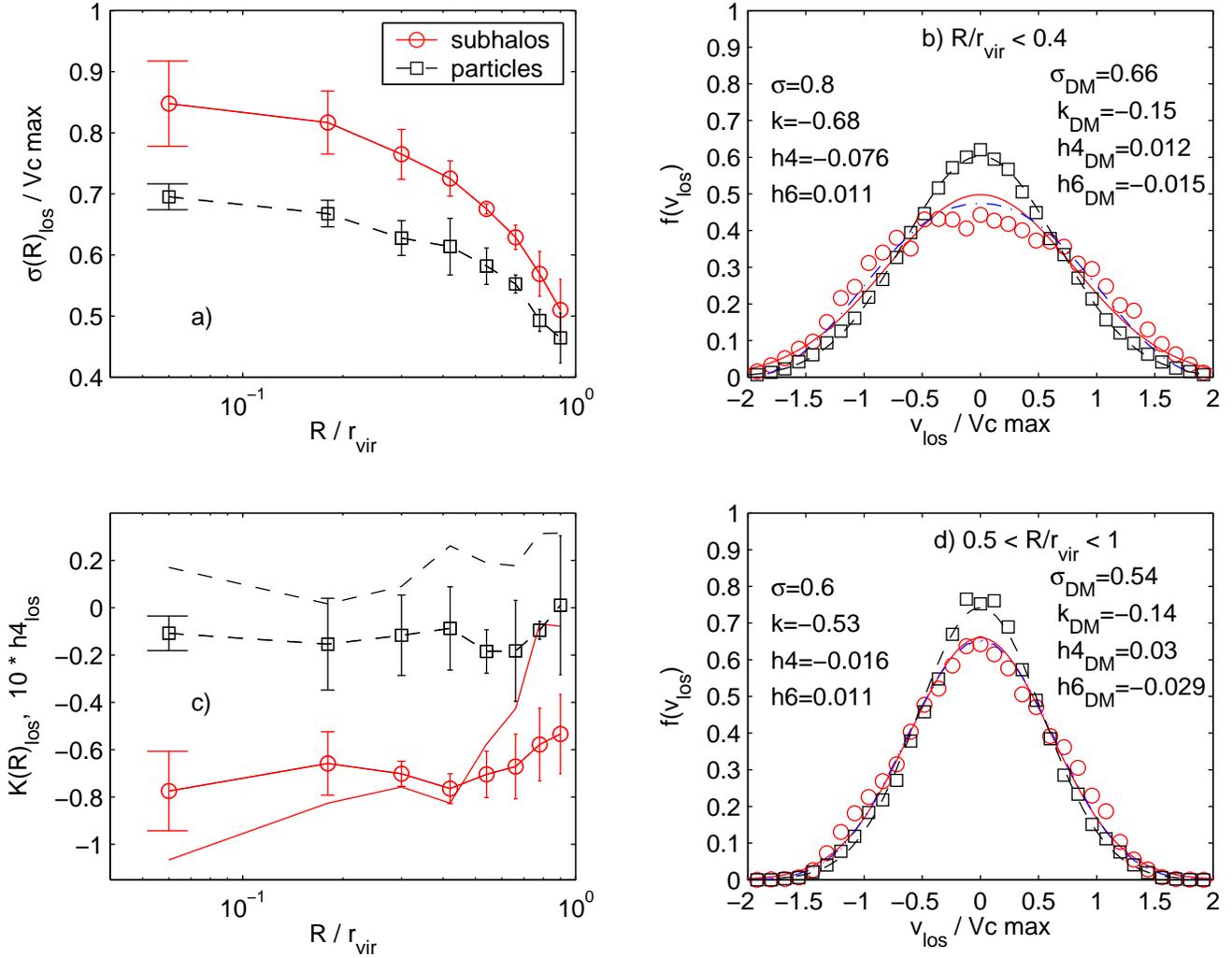}
\caption{\label{biasProj.eps}Panel a): Average line of sight velocity
dispersion of subhalos and particles as a function of projected
distance from the center. Panel b) and panel
d): Average line of sight velocity distributions of subhalos and
particles, for projected radii smaller (b) and larger (d) than 0.4
$r_{\rm virial}$. Solid and dashed lines are Gaussians with a
second moments fitted to the subhalos (solid) and to the particles
(dashed). Fourth order Gauss-Hermite approximations to the subhalo velocity
distribution functions are given with dashed-dotted lines.  
Panel c): Average kurtosis (with error bars) and fourth Gauss-Hermite moment 
(without error bars and multiplied by a factor of ten for clarity) 
of the line of sight velocity
components of subhalos and particles as a function of projected
distance from the center. The error bars in panels a) and c) give the scatter within 
the six clusters.}
\end{figure*}

Figure \ref{biasProj.eps} shows the projected moments of the CDM subhalo 
velocity distributions and the inner and outer distribution of line of sight velocities
averaged. The ploted values are averages over the six cluster halos and over
three different projections.
The velocity moments for the dark matter background are also plotted 
for comparison, a similar analysis was presented by \citet{Sanchis2004}.

In contrast to the spatial distribution the {\it velocity} 
distribution of CDM subhalos agrees surprisingly well with
current observations of cluster galaxies. In the grand total 
velocity distribution of the CNOC survey a negative $h_4 = -0.015 \pm 0.005$ was found
\citep{VanDerMarel2000} and $h_6 = -0.028 \pm 0.006$. We get $k =
-0.44$, $h4 = -0.016$ and $h_6 = -0.022$ using all subhalos with bound
mass larger than $5\times10^{-5} M_{\rm virial}$. There are 1152 subhalos
above this threshold in our six clusters. 
This agreement between simulations and observations may be 
fortuitous since the spatial distribution of galaxies is different
and probably due to destruction of low energy central subhalos.
Also, in the Coma cluster
the velocity distribution seems to be more flat topped compared to a
Gaussian: The kurtosis is negative in most radial bins, the values
scatter around $k\simeq -0.5$ (see Figure 3 in \cite{Lokas2003}). The
uncertainties in the measurement of velocity moment profiles are still
quite large and a comparison with the projected moments from Figure
\ref{biasProj.eps} of the CDM subhalo velocities is not feasible yet.

\section{Conclusions}\label{Conclusions}

We analyse the substructure within six very high resolution 
cold dark matter simulations of galaxy clusters 
and four simulations of galaxies.
We have addressed several open issues raised in the introduction
regarding the results of high resolution simulations of individual
halos within the concordance CDM model. Our conclusions can be
summarised as follows:

\begin{enumerate}
\item The spatial distribution of subhalos in cold dark matter
simulations of galaxies and clusters is antibiased with respect to the
mass. Although this behavior was found by other groups, we demonstrate
that this result is robust and
does not change as we increase the resolution. We show that this antibias
most likely results from a population of early halos that are
tidally destroyed in the dense protocluster environment and within the 
central regions of the final cluster.
\item The surviving population of subhalos have a positive velocity
bias that increases towards the center of the halos.
The subhalo velocity distribution functions are non-Gaussian,
they are 'flat topped', especially in the inner region:
There the kurtosis is $k=-0.7$
and the fourth Gauss-Hermite coefficient $h_4 = -0.068$.
\item The spatial anti-bias and the positive velocity bias of the subhalos 
are consistent with a steady-state solution of the Jeans equation.
Subhalos are a hot, more extended component in equilibrium with the
potential generated by the smooth particle background.
\item The mass and circular velocity distributions of subhalos in
our highest resolution simulation show the same power law slopes as 
in lower resolution versions, but are steeper at the low mass end.
It is not clear that
convergence in the number of subhalos has been reached
below a scale of a few hundred particle masses.
\item Cluster and galaxy mass halos simulated at the
same resolution have similar substructure abundances.
The scatter in the circular velocity and mass functions is a
factor of three
at the high mass end, but falls to just 1.7 at lower masses. 
\item An observational comparison with CNOC cluster data and the 
Coma cluster shows
that the galaxy population traces the smooth dark matter background,
but not the predicted halo population. 
This is most likely due to overmerging
in the central region of the simulations and we are probably missing a
factor of two in the subhalo population. The baryonic cores of
these disrupted subhalos may survive intact if dissipational
processes increase their densities sufficiently. Also a 
greatly truncated dark
matter halo may survive in this case.

\end{enumerate}

This latter statement is the most profound conclusion of this work.
The spatial distribution of cluster galaxies is significantly 
different from the distribution of subhalos in dark matter simulations. 
Either the model is incorrect or we have reached a fundamental limit to 
this type of pure dark matter simulation. 
Here we explore the latter possibility and the
implications for the morphology density relation.

It is likely that disk galaxies do not significantly modify the overall
potential provided by the baryons and dark matter. Whereas a disk-disk
merger would funnel gas to the central region, forming an elliptical
galaxy with a significantly deeper potential and a effective rotation
curve that is at least isothermal, or possibly 
Keplerian in the center \citep{Rom2003}.  
Thus we expect that an
elliptical galaxy would most likely survive at any position within the
cluster, albeit with a greatly truncated dark matter halo. Late type 
spiral galaxies are unlikely to survive within the central regions of
clusters (or their progenitors) and will become physically overmerged 
to form the cD halo of diffuse light. 

If the CDM paradigm is correct then we are missing close to a factor
of two of the `galaxy' population as associated with subhalos, increasing
to a factor of five within the inner 10\% of the cluster. It is
possible that simulations with more than $10^9$ particles per system
may resolve more central subhalos and calculations this large will be
possible in the future. In this case, the velocity 
bias should decrease as we
resolve more halos/galaxies in the central regions.  However, from our
convergence study we find very few new halos in the central cluster
regions as we increase the resolution by a factor of ten.
This implies that we have reached a physical limit to
DM-only simulations and that any loss of subhalos in current
simulations is due to physical overmerging 
(\citealt{White1978}; \citealt{Moore1996}). In this case progress in
this area can only be made by including a realistic treatment of
hydrodynamics and star-formation such that realistic disks and
elliptical galaxies can be followed within the appropriate 
cosmological context.

The survival or disruption of a galaxy depends on an intricate balance 
between the progenitors dark halo 
structure and the effects of dissipation. Sa-Sb
galaxies must lie on the borderline between survival and disruption
in the cluster environment.
The morphology-density relation
may simply reflect the fact that
the disks are preferentially destroyed in the central regions of clusters.
However if the CDM model is correct one needs to
preferentially form ellipticals in high density regions before the
cluster forms. The fact that the observed galaxy distribution follows
the dark matter distribution implies that no overmerging of galaxies
has taken place. It is insufficient to take disks and destroy them in
the cluster cores since this would give rise to a cored galaxy
distribution.

The fact that 40\% of halos identified at z=7 can not be associated
with a subhalo at z=0, or have not merged with the central cD, implies that
they have merged into the smooth particle background. If these objects
can be associated with surviving galaxies, it implies a strong age-radius
dependence for galaxies within clusters. At the cluster centres
over 80\% of the galaxies must have formed prior to z=7.

\section*{Acknowledgments}

We thank the referee for many insightful comments and
suggestions.
We are grateful to Ewa Lokas for kindly providing the galaxy number density
data for the Coma cluster and to Frank van den Bosch,
Chiara Mastropietro, Peder Norberg and Jeremiah Ostriker for useful discussions.  
We thank the Swiss Center for Scientific
Computing in Manno for computing time, we generated the initial
conditions there. The simulations were performed on the zBox
\footnote{http://www-theorie.physik.unizh.ch/$\sim$stadel/zBox/}
supercomputer at the University of Zurich.  J. D. is supported by the
Swiss National Science Foundation.

\bsp
\label{lastpage}

\begin{thebibliography}{99}

\bibitem[\protect\citeauthoryear{Bertschinger}{2001}]{Bertschinger2001}
Bertschinger E., 2001, ApJSS, 137, 1 

\bibitem[\protect\citeauthoryear{Binney \& Tremaine}{1987}]{Binney1987}
Binney J., Tremaine S., 1987, Galactic Dynamics.
Princeton Univ. Press, Princeton

\bibitem[\protect\citeauthoryear{Beers \& Tonry}{1986}]{Beers1986}
Beers T.~C.~, Tonry J.~L., 1986, ApJ, 300, 557 

\bibitem[\protect\citeauthoryear{Carlberg \& Couchman}{1989}]{Carlberg1989}
Carlberg R.~G., Couchman H.~M.~P., 
1989, ApJ, 340, 47 

\bibitem[\protect\citeauthoryear{Carlberg}{1994}]{Carlberg1994}
Carlberg, R.~G., 1994, ApJ, 433, 468

\bibitem[\protect\citeauthoryear{Carlberg, Yee \& ~Ellingson}{Carlberg et al.}{1997}]{Carlberg1997}
Carlberg R. G., Yee H. K. C., Ellingson E., 1997, ApJ, 478, 462

\bibitem[\protect\citeauthoryear{Colin, Klypin, \& Kravtsov}{2000}]{Colin2000}
Colin P., Klypin A.~A., Kravtsov A.~V., 2000, ApJ, 539, 561 

\bibitem[\protect\citeauthoryear{Cypriano et al.}{2001}]{Cypriano2001}
Cypriano E.~S., Sodr{\' e} L.~J., Campusano L.~E., Kneib J., Giovanelli R., Haynes M.~P., Dale D.~A., Hardy 
E., 2001, AJ, 121, 10 

\bibitem[\protect\citeauthoryear{Davis et al.}{1985}]{Davis1985}
Davis M., Efstathiou G., Frenk C. S., White S.D.M., 1985, ApJ, 292, 371

\bibitem[\protect\citeauthoryear{Desai et al.}{2004}]{Desai2004}
Desai V., Dalcanton J. J., Mayer L., Reed D., Quinn T., Governato F., 2004, preprint
, astro-ph/0311511

\bibitem[\protect\citeauthoryear{De Lucia et al.}{2004}]{DeLucia2004}
De Lucia G., Kauffmann G., Springel V., 
White S.~D.~M., Lanzoni B., Stoehr F., Tormen G., Yoshida N., 2004, MNRAS, 
348, 333 

\bibitem[\protect\citeauthoryear{Diemand et al.}{2004a}]{Diemand2004a}
Diemand J., Moore B., Stadel J., Kazantzidis S., 2004a, MNRAS, 348, 977 

\bibitem[\protect\citeauthoryear{Diemand, Moore \& Stadel}{Diemand et al.}{2004b}]{Diemand2004b}
Diemand J., Moore B., Stadel J., 2004b, preprint, astro-ph/0402267

\bibitem[\protect\citeauthoryear{Frenk et al.}{1996}]{Frenk1996}
Frenk, C.~S., Evrard  A.~E.~, White  S.~D.~M., Summers F.~J.,1996, ApJ, 472, 460

\bibitem[\protect\citeauthoryear{Gerhard}{1993}]{Gerhard1993} 
Gerhard O.~E., 1993, MNRAS, 265, 213 

\bibitem[\protect\citeauthoryear{Ghigna et al.}{1998}]{Ghigna1998} 
Ghigna S., Moore B., Governato F., Lake G., Quinn T., Stadel J., 1998, 
MNRAS, 300, 146 

\bibitem[\protect\citeauthoryear{Ghigna et al.}{2000}]{Ghigna2000}
Ghigna S., Moore B., Governato F., Lake G., Quinn T., Stadel J., 2000, ApJ, 544, 616

\bibitem[\protect\citeauthoryear{Jenkins et al.}{2001}]{Jenkins2001}
Jenkins A., Frenk C.~S., White S.~D.~M., 
Colberg J.~M., Cole S., Evrard A.~E., Couchman H.~M.~P., Yoshida N., 2001, 
MNRAS, 321, 372 

\bibitem[\protect\citeauthoryear{Kazantzidis et al.}{2004}]{Kaz2004}
Kazantzidis S., Mayer L., Mastropietro C., Diemand J., Stadel J., Moore B., 2004, 
ApJ in press, astro-ph/0312194

\bibitem[\protect\citeauthoryear{Klypin et al.}{1999a}]{Klypin1999a} 
Klypin A., Gottl{\" o}ber S., Kravtsov A.~V., Khokhlov A.~M., 1999, ApJ, 
516, 530 

\bibitem[\protect\citeauthoryear{Klypin et al.}{1999b}]{Klypin1999b} 
Klypin A., Kravtsov A.~V., Valenzuela O., Prada F., 1999, ApJ, 522, 82 

\bibitem[\protect\citeauthoryear{Kravtsov, Gnedin \& Klypin}{Kravtsov et al.}{2004}]{Kravtsov2004}
Kravtsov A.~V.~, Gnedin O.~Y., Klypin A.~A., 2004, ApJ in press, astro-ph/0401088

\bibitem[\protect\citeauthoryear{Lokas \& Mamon}{2003}]{Lokas2003}
Lokas E.~L., Mamon G.~A., 2003, MNRAS, 343, 401 

\bibitem[\protect\citeauthoryear{Merrifield \& Ken}{1989}]{Merrifield1989} 
Merrifield M.~R.~, Kent S.~M., 1989, AJ, 98, 351 

\bibitem[\protect\citeauthoryear{Moore, Katz \& Lake}{Moore et al.}{1996}]{Moore1996}
Moore B., Katz N., Lake G., 1996, ApJ, 457, 455

\bibitem[\protect\citeauthoryear{Moore et al.}{1998}]{Moore1998}
Moore B., Governato F., Quinn T., Stadel J., Lake G., 1998, ApJ, 499, L5

\bibitem[\protect\citeauthoryear{Moore et al.}{1999}]{Moore1999} 
Moore B., Ghigna S., Governato F., Lake G., Quinn T., Stadel J., Tozzi P., 
1999, ApJ, 524, L19 

\bibitem[\protect\citeauthoryear{Navarro, Frenk \& White}{Navarro et al.}{1996}]{Navarro1996}
Navarro J. F., Frenk C. S., White S. D. M., 1996, ApJ, 462, 563

\bibitem[\protect\citeauthoryear{Okamoto \& Habe}{1999}]{Okamoto1999}
Okamoto T., Habe A., 1999, ApJ, 516, 591 

\bibitem[\protect\citeauthoryear{Reed et al.}{2003}]{Reed2003} 
Reed D., Gardner J., Quinn T., Stadel J., Fardal M., Lake G., Governato F., 
2003, MNRAS, 346, 565 

\bibitem[\protect\citeauthoryear{Romanowsky et al.}{2003}]{Rom2003}
Romanowsky A.~J., Douglas N.~G., Arnaboldi 
M., Kuijken K., Merrifield M.~R., Napolitano N.~R., Capaccioli M., Freeman 
K.~C., 2003, Sci, 301, 1696 

\bibitem[\protect\citeauthoryear{Sanchis, Lokas \& Mamon}{2004}]{Sanchis2004}
Sanchis T., Lokas E. L., Mamon G. A., 2004, MNRAS, 347, 1198

\bibitem[\protect\citeauthoryear{Spergel et al.}{2003}]{Spergel2003}
Spergel D. N. et al., 2003, ApJSS, 148, 175

\bibitem[\protect\citeauthoryear{Springel et al.}{2001}]{Springel2001} 
Springel V., White S.~D.~M., Tormen G., Kauffmann G., 2001, MNRAS, 328, 726 

\bibitem[\protect\citeauthoryear{Stadel}{2001}]{Stadel2001}
Stadel J., 2001, PhD thesis, U. Washington

\bibitem[\protect\citeauthoryear{Stoehr et. al}{2002}]{Stoehr2002}
Stoehr F., White S. D., Tormen G., Springel V., 2002, MNRAS, 335, L84

\bibitem[\protect\citeauthoryear{Summers, Davis \& Evrard}{Summers et al.}{1995}]{Summers1995}
Summers, F.~J., M.~Davis, A.~E.~Evrard 1995, ApJ, 454, 1

\bibitem[\protect\citeauthoryear{Taffoni et al.}{2003}]{Taffoni2003}
Taffoni G., Mayer L., Colpi M., Governato F., 2003, MNRAS, 341, 434

\bibitem[\protect\citeauthoryear{Taylor, Silk, \& Babul}{Taylor}{2003}]{Taylor2003}
Taylor J.~E., Silk J., Babul A., 2003, IAUS, 220, astro-ph/0312086

\bibitem[\protect\citeauthoryear{Tormen, Bouchet \& White}{Tormen}{1997}]{Tormen1997}
Tormen G., Bouchet F.~R., White 
S.~D.~M., 1997, MNRAS, 286, 865 

\bibitem[\protect\citeauthoryear{van der Marel et al.}{2000}]{VanDerMarel2000}
van der Marel R.~P., Magorrian J., 
Carlberg R.~G., Yee H.~K.~C., Ellingson E., 2000, AJ, 119, 2038 

\bibitem[\protect\citeauthoryear{van Kampen}{1995}]{vanKampen1995}
van Kampen E.,1995, MNRAS, 273, 295

\bibitem[\protect\citeauthoryear{White}{1976}]{White1976}
White, S.~D.~M., 1976, MNRAS, 177, 717 

\bibitem[\protect\citeauthoryear{White \& Rees}{1978}]{White1978}
White, S.~D.~M., Rees M.~J., 1978, MNRAS, 183, 341

\bibitem[\protect\citeauthoryear{White et al.}{1987}]{White1987}
White, S.~D.~M., M.~Davis, G.~Efstathiou, and C.~S.~Frenk, 1987, Nature, 330, 45

\end{thebibliography}
\end{document}